\documentclass[aps,prl,twocolumn]{revtex4}

\usepackage{amssymb,amsmath,graphicx,graphics,color,epstopdf}

\usepackage{dblfloatfix}
\usepackage{float}

\newcommand{\ket}[1]{|#1\rangle}
\newcommand{\bra}[1]{\langle#1|}

\newcommand{\braket}[2]{\langle#1|#2\rangle}

\begin{document}

\title{Monopole Antimonopole Instability in Non-Hermitian Coupled Waveguides}
\author{Rosie Hayward}
\author{Fabio Biancalana}
\affiliation{School of Engineering and Physical Sciences, Heriot-Watt University, EH14 4AS Edinburgh, UK}

\begin{abstract}
A non-Hermitian coupled waveguide system with periodically varying parameters, in which the Berry curvature is analogous to a hyperbolic magnetic monopole or antimonopole, is investigated. It is shown to have a purely imaginary Berry connection, and is consequently influenced by a geometric multiplier. It is possible for this multiplier to induce net gain or loss in the system, corresponding to the existence of the antimonopole or monopole in parameter space, respectively. For the right choice of parameters, the system will display an apparent non-adiabatic change in behaviour, which implies a switch between the dominant eigenstate in the waveguides, leading to a change in parameter space analogous to a charge reversal of the hyperbolic magnetic monopole.

\end{abstract}

\maketitle

\textit{Introduction.}--- When the external parameters of a quantum mechanical system change slowly, through making the adiabatic approximation, we can induce a gauge potential, the Berry connection, and hence a gauge field, the Berry curvature, which will give rise to Berry's geometric phase \cite{Berry}. All arise from the geometric properties of the system's eigenfunctions in the space of its varying parameters. The geometric phase is found by integrating the connection over a closed loop in parameter space, and is akin to the Aharonov-Bohm phase of a charged particle traversing a loop containing magnetic flux; the curvature in this case is like a magnetic field \cite{Berry, Aharonov, Xiao}. The analogy between the curvature field and magnetic fields is well known \cite{Price, Wu}, and can be extended to magnetic monopoles by the fact that the integral of the curvature over closed surfaces is known to be topological and of integer value, as would be the case for Dirac monopoles with quantised magnetic charge \cite{Xiao}. 

Magnetic monopoles remain elusive in real space, due to their predicted very heavy mass \cite{Milton}. However, they have been observed in dual space, i.e. the momentum space of solids, via the Berry connection of Bloch wavefunctions, which in this case will give rise to the transverse conductivity when integrated over a closed loop \cite{Fang}. There are many theoretical examples where the magnetic monopole corresponds to the source or sink of the Berry curvature, mainly in condensed matter physics \cite{Nesterov, Haldane, Zhang, Murakami}. Note that these examples are distinct from the condensed matter spin-ice emergent magnetic monopoles \cite{Mengotti}.

The `magnetic monopole' proposed here exists not in $k$-space, or real space, but in the space of the varying optical parameters of a non-Hermitian, $\mathcal{PT}$-symmetric coupled waveguide system, similar to those seen in \cite{RosieandFabio, Rotter, Guo, Ramezani, El-Ganainy, Ruter}. In such a system, it is possible for the geometric phase to become complex, or even purely imaginary, and hence no longer represent a true phase \cite{GandW, Berry2}. In this case, we find a hyperbolic monopole and antimonopole gauge field associated with each instantaneous eigenstate of the system, as seen in \cite{Nesterov}, which will each have an associated imaginary flux. This will lead to a real `geometric multiplier' as seen in \cite{RosieandFabio}, which can be non-zero for a closed loop in parameter space due to the non-trivial geometry, and will in turn affect the system dynamics beyond modulation, potentially inducing an interesting interplay between gain and loss.  

We show that analytical calculations of the monopole flux through a loop can be used to predict the gain and loss seen in simulations to a high degree of accuracy, showing that the existence of the monopole (or antimonopole) in parameter space indeed induces an instability in the system. This is an example of a geometric instability, due to its dependence on the non-Hermitian Berry connection. If one does not suppress the non-adiabatic evolution of the system, it is possible to induce an apparent eigenstate-flip, signalling a charge reversal of the monopole in parameter space. 

\begin{figure}[h]
\centering
\includegraphics[width=0.5\textwidth]{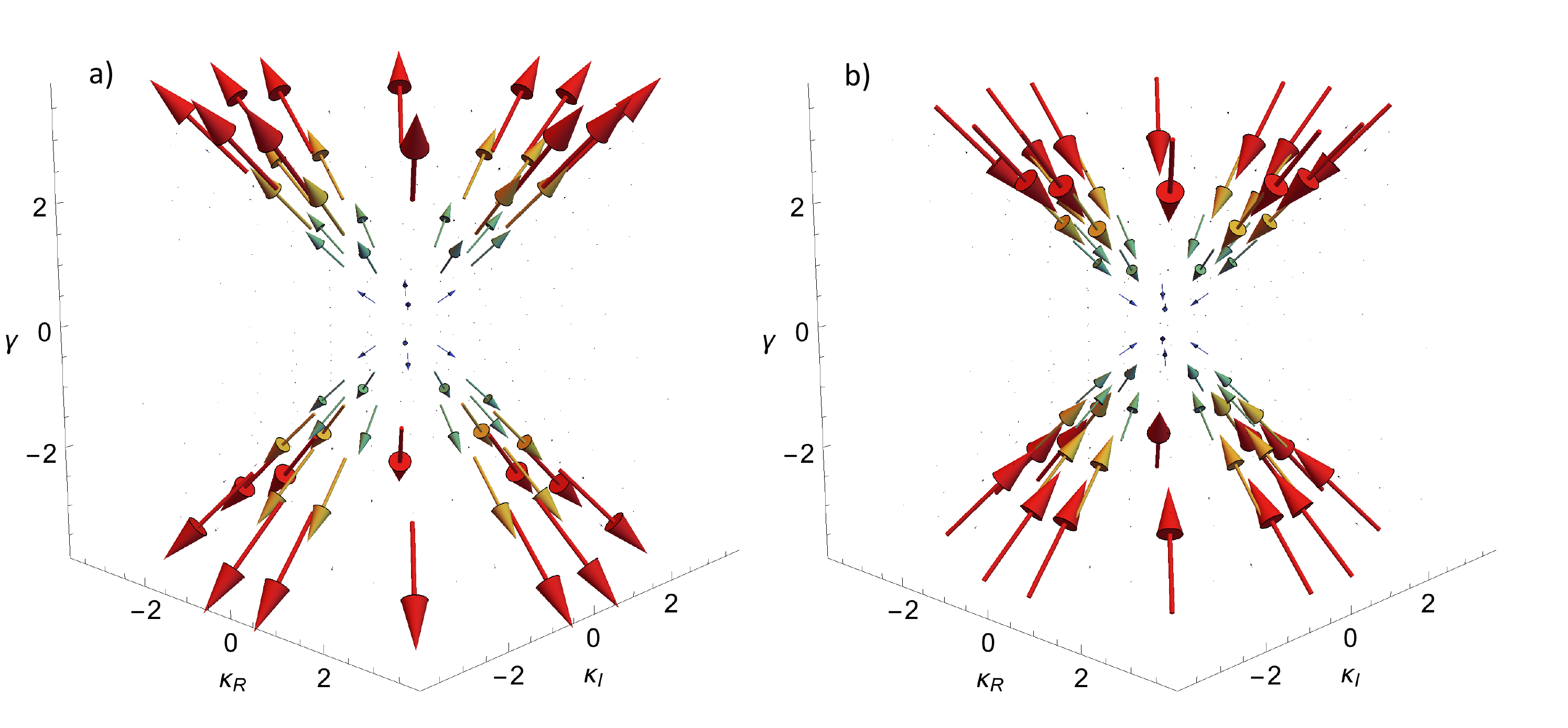}
\caption{a) The monopole $( \Omega^+)$ field, and b) the anti-monopole field  $( \Omega^-)$. In both cases, the imaginary unit seen in Equation (\ref{bfield}) has been neglected. Although the field lines do flow out of (into) a point, the divergence is of course still zero, due to the fact the field is found by taking the curl of a potential.   }
\label{field}
\end{figure}

\textit{The hyperbolic pseudo-magnetic monopole.}--- Consider a $\mathcal{PT}$-symmetric coupled waveguide system with balanced gain and loss, and a complex coupling coefficient, both which vary periodically along the length of the waveguides. Such systems can be experimentally realised via the fabrication of waveguides with sinusoidal modulation along their length \cite{Eichelkraut}. Use of such a technique is experimentally demonstrated in \cite{Rotter}, where coupling and detuning coefficients are varied to form a closed loop in parameter space. Our coupled waveguides can be described by a non-Hermitian Hamiltonian, akin to the one physically realised in \cite{Rotter}, and also \cite{Guo}, where there is no parameter variation but a complex coupling coefficient is considered; complex couplings are also found in distributed feedback lasers, which can be modelled as coupled waveguides \cite{complexcoupling}. Hence, the following Hamiltonian represents a general, realistic physical coupled waveguide system:
\begin{equation}\label{eq:ham}
\hat{H}(z)=\begin{pmatrix} \, i \, \gamma(z) && \kappa(z) \\ \kappa^*(z) && -i \,\gamma(z) \end{pmatrix}.
\end{equation}

\begin{figure}[h]
\centering
\includegraphics[width=0.4\textwidth]{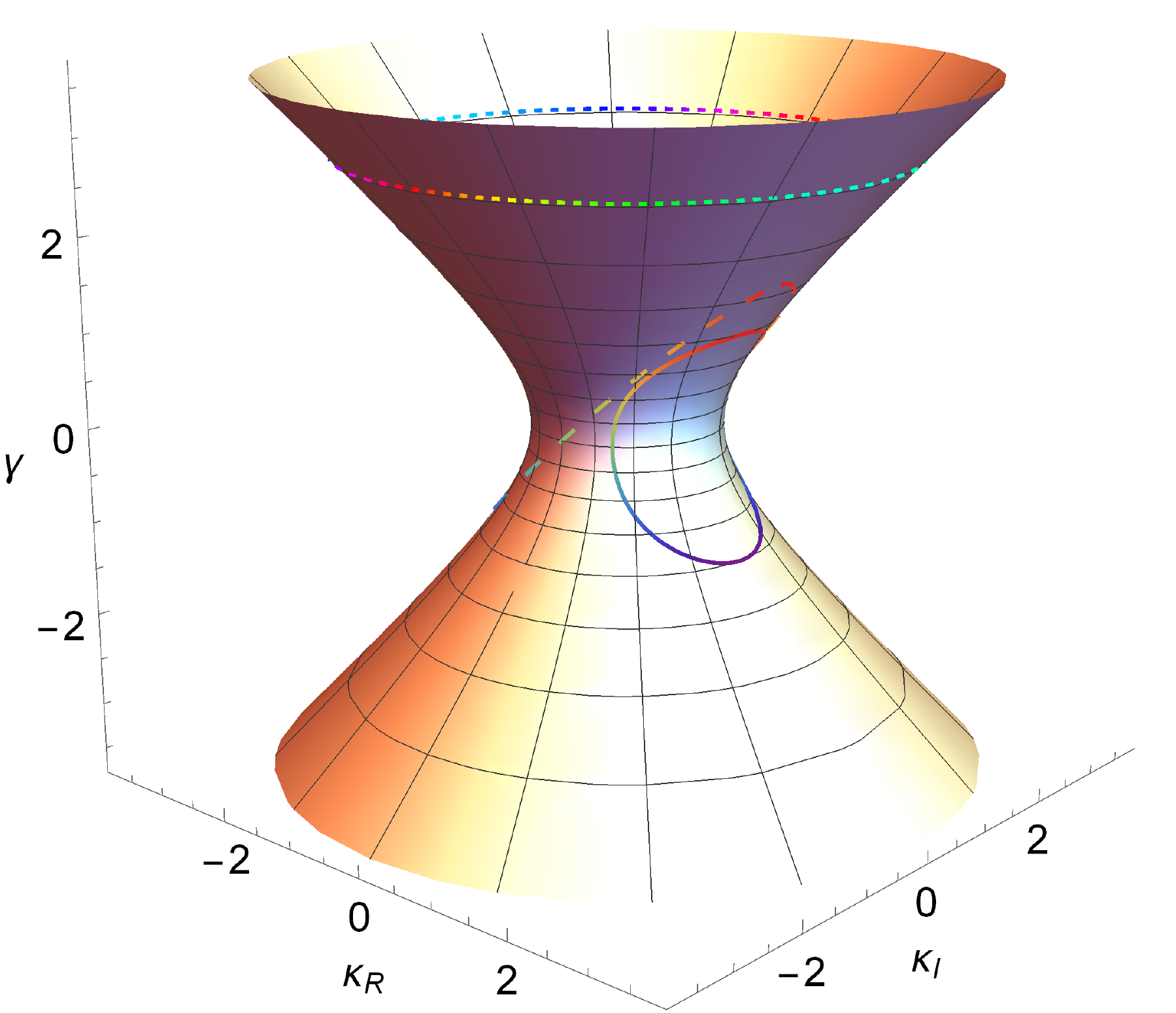}
\caption{Example loops on the surface of constant flux associated with the monopole and antimonopole. The tilted loop around the neck of the surface will yield zero net flux, and hence the system will remain stable as its radius increases. For the other loops, the net flux, and consequently $\gamma_B$, will be non-zero for a complete cycle, causing the system to exhibit gain or loss. }
\label{figloops}
\end{figure}

Here, $\kappa(z)=\kappa_R(z) +i \kappa_I(z)$ is the complex coupling coefficient, $\gamma(z)$ represents a gain/loss coefficient, and $z$ is the distance along the length of the waveguides. The form of the above Hamiltonian is derived from that seen in \cite{Nesterov}. We can, for simplicity, substitute $\gamma(z)=r(z)\sin(\alpha(z))$ and $\kappa_R(z)+i \kappa_I(z)=r(z)e^{i \theta(z)}$ in (1), which gives rise to the instantaneous eigenstates,
\begin{equation}\label{eig}
\ket{\psi_+}=\frac{1}{\sqrt{2}}\begin{pmatrix} e^{i \frac{\alpha+\theta}{2}} \\ e^{-i \frac{\alpha+\theta}{2}} \end{pmatrix}, \, \ket{\psi_-}=\frac{1}{\sqrt{2}}\begin{pmatrix} - \, e^{-i \frac{\alpha-\theta}{2}} \\  \, e^{i \frac{\alpha-\theta}{2}} \end{pmatrix},
\end{equation}
which correspond to the instantaneous eigenvalues $\lambda_{\pm}=\pm\sqrt{x^2+y^2-\gamma^2}=\pm\lambda$, the form of which already hints at a hyperbolic geometric structure of the parameter space. We can also find the adjoint-eigenstates by solving for the eigensystem of $\hat{H}^{\dagger}$:
\begin{equation}\label{adj}
\ket{\phi_+}=\frac{1}{\sqrt{2}}\begin{pmatrix} e^{-i \frac{\alpha-\theta}{2}} \\ e^{i \frac{\alpha-\theta}{2}} \end{pmatrix}, \, \ket{\phi_-}=\frac{1}{\sqrt{2}}\begin{pmatrix}   -e^{i \frac{\alpha+\theta}{2}} \\   e^{-i \frac{\alpha+\theta}{2}} \end{pmatrix},
\end{equation}
which again correspond to $\lambda_{\pm}$ respectively, as long as the system remains in the unbroken $\mathcal{PT}$-symmetry phase ($\lambda^2>0$).

From our instantaneous eigenstates and adjoint-eigenstates, we can find the non-Hermitian Berry connection, and the associated pseudo-magnetic field found in the space of our varying parameters. As we are working with a non-Hermitian system, we define the connection ${\bf A}^{\pm}({\bf R})=(A_x^{\pm}, A_y^{\pm}, A_{\gamma}^{\pm})$ for ${\bf R}=(\kappa_R, \kappa_I, \gamma)$ as follows, in line with the standard definition for the geometric phase in this case \cite{GandW, Berry2}:
\begin{equation}\label{connection}
A_\mu^{\pm}=i\frac{\bra{\phi_\pm}\partial_\mu\ket{\psi_\pm}}{\braket{\phi_\pm}{\psi_\pm}},
\end{equation}
where the index $\mu=\{\kappa_R,\kappa_I,\gamma\}$. Inserting (\ref{eig}) and (\ref{adj}) into (\ref{connection}), and taking the curl of the connection with respect to the system's parameters, in line with the standard definition of the Berry curvature, gives us the following induced pseudo-magnetic fields:

\begin{equation} \label{bfield}
\mathbf{\Omega}^{\pm}=\frac{\pm i}{2\lambda^3}\begin{pmatrix} \kappa_R \\ \kappa_I \\ \gamma \end{pmatrix}, 
\end{equation}
which correspond to a source (+) and a sink ($-$), with a `magnetic charge' of $\pm \frac{1}{2}$.

Figure \ref{field} shows the pseudo-magnetic vector fields of the monopole and antimonopole (the imaginary unit has been neglected), plotted in the space of the system's varying parameters. As with ordinary magnetic monopoles, all the field lines emanate from a point, but the hyperbolic monopole is unusual in the way the field strength increases with distance from the source.

\textit{The geometric multiplier.}---  For a cyclic Hamiltonian, if the adiabatic approximation is upheld, then eigenstates will gain a geometric phase once the system's parameters have returned to their original values, and otherwise remain unchanged. In non-Hermitian systems, it is no longer a requirement that Berry's geometric phase is a real function, and in general it is complex, no longer representing a true phase \cite{GandW}. In terms of (\ref{connection}), we find explicitly the geometric function:
\begin{equation}\label{geometric}
\gamma_B^{\pm}(z)=\int_0^z {\bf A}^{\pm}({\bf R}) \, \mathrm{d} {\bf R} ,
\end{equation}
\begin{equation}\label{geometric1}
=\int_0^z -\frac{i}{2} \tan{[\alpha(z')]} (\alpha'(z')\pm\kappa'(z')) \, d z' .
\end{equation}

\begin{figure}[h]
\centering
\includegraphics[width=0.45\textwidth]{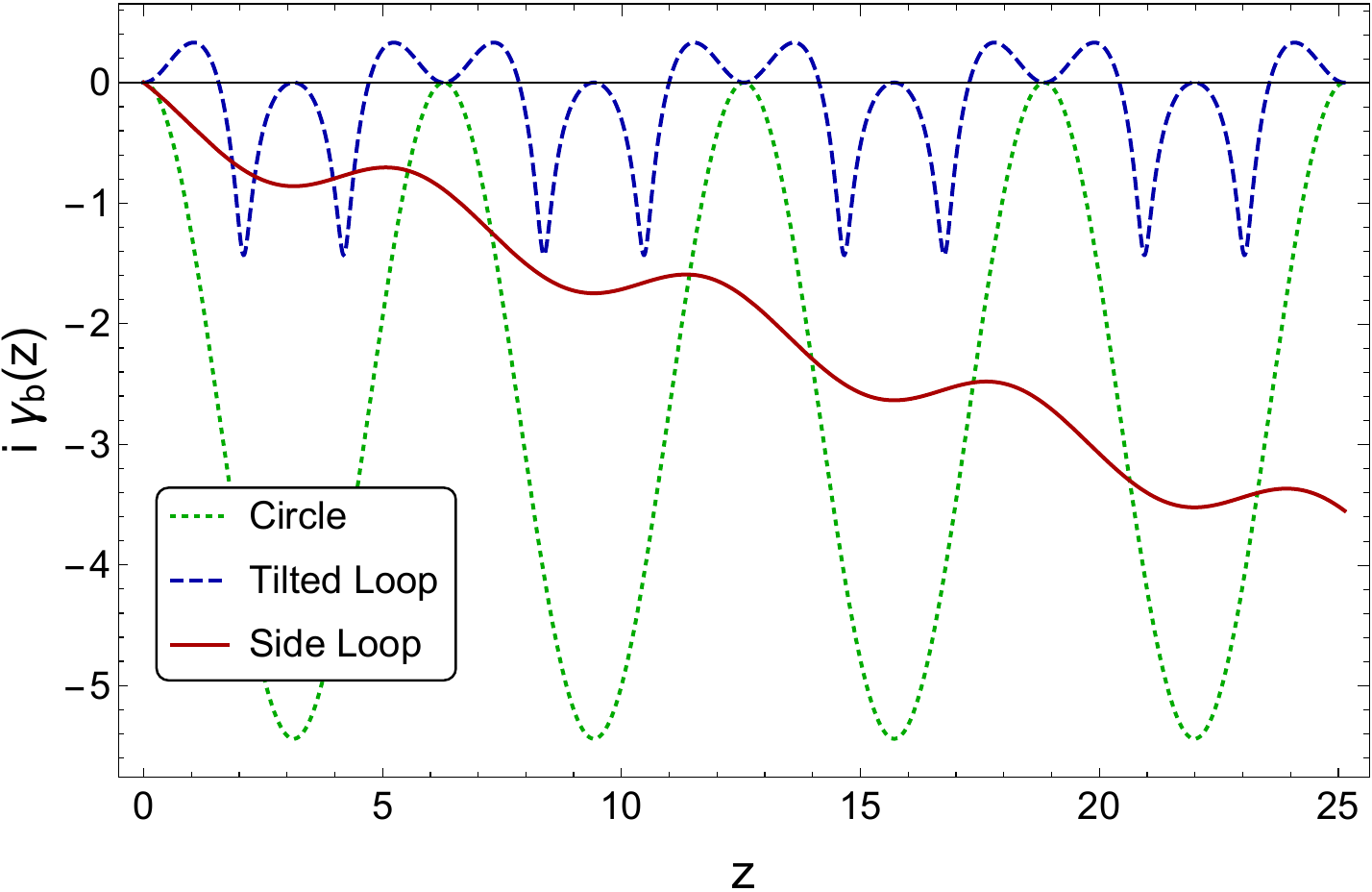}
\caption{The variation of the geometric function $i \gamma_B(z)$ with the distance along the coupled waveguides, $z$, for a circular loop $(a=2, b=1)$, a tilted loop $(a=3, b=1)$, and a side loop $(a=1 ,b=0.5)$, as seen in Figure \ref{figloops}.  }
\label{berry}
\end{figure}

In this case, as the connection itself is imaginary, eigenstates will hence be influenced by a real geometric multiplier \cite{RosieandFabio}. As this multiplier is a periodic function, even if it returns to zero after a complete cycle, it will magnify the non-Hermitian analogue of Floquet sidebands in the spectrum of the coupled waveguide system. Furthermore, it can be expected to modulate the amplitude of the instantaneous eigenstates as they evolve along the waveguides. If the geometric multiplier is non-periodic, we can expect it to introduce either net gain or loss into our system by changing the amplitude of the waveguide modes by a fixed amount every time a cycle is completed. 

The energy surface ($\lambda^2={\rm constant}$) which corresponds to our eigenstates in parameter space, takes the form of a one-sheet hyperboloid for $\lambda^2>0$. We are free to choose our parameters such that $\lambda^2=1$; this is automatically true when $\kappa_R=\cosh{\eta}\cos{\theta}$, $\kappa_I=\cosh{\eta}\sin{\theta}$, and $\gamma=\sinh{\eta}$. Hence, we can find closed, cyclic paths on our hyperboloid's surface which will correspond to periodic variations in our waveguide parameters by giving $\theta$ and $\eta$ a periodic variation in $z$. Examples of possible closed paths corresponding to cycles of the Hamiltonian the parameters could take on the energy surface, and which can consequently be used to calculate the corresponding geometric phase, are shown in Figure \ref{figloops}.

\begin{figure}[h]
 \includegraphics[width=0.5\textwidth]{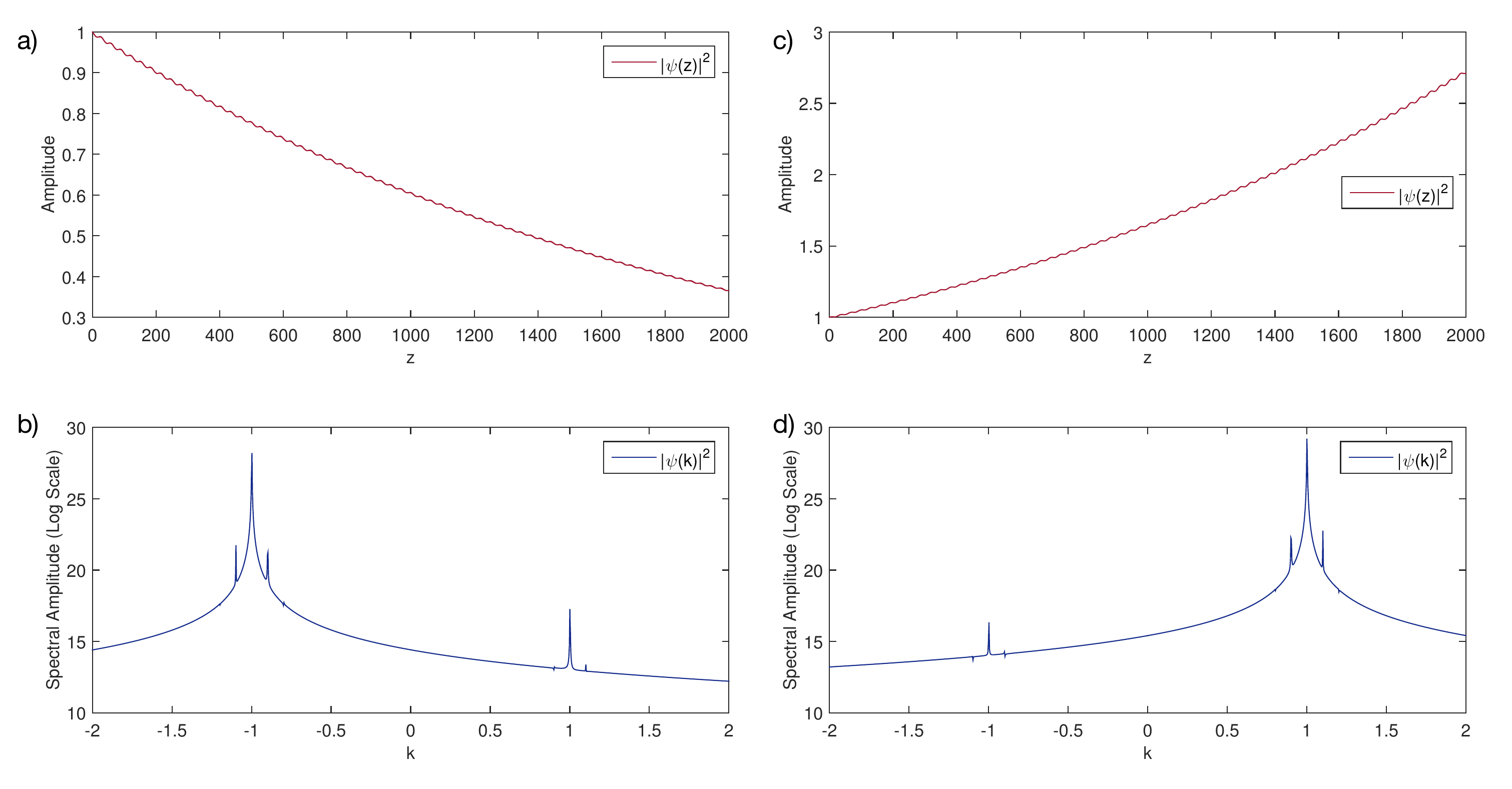}
\caption{a) Net loss of a wavefunction evolving along the coupled waveguide system for $a=0.1$, $b=0.1$, when the initial instantaneous eigenstate is $\ket{\psi_+}$, due to the presence of the geometric multiplier $e^{i \gamma_B}$. b) The spectrum corresponding to a). c) Net gain of a wavefunction evolving along the coupled waveguide system for $a=0.1$, $b=0.1$, when the initial instantaneous eigenstate is $\ket{\psi_-}$, due to the presence of the geometric multiplier $e^{i \gamma_B}$. d) The spectrum corresponding to c). }
\label{gainandloss}
\end{figure}

Due to the correspondence between the Berry connection (\ref{connection}) and the magnetic vector potential, the integral above is identical to the integral one would use to calculate the flux of the hyperbolic monopole or antimonopole through a closed loop, such as those seen in Figure \ref{figloops}. Hence, we can relate the net flux out of (in to) the monopole (antimonopole) to the presence of gain or loss in our system, given the evolution is suitably adiabatic that other effects can be neglected. One can calculate an approximate ratio of amplitude of the wavefunction at the end of the waveguides ($|\psi(L)|^2$), to the amplitude of the initially excited eigenstate ($|\psi_{\pm}|^2$), due to the presence of a non-periodic geometric multiplier from the following, by assuming the wavefunction gains a factor $e^{i \gamma_B^{\pm} }$ at the end of each cycle:

\begin{equation} \label{multiplier}
\frac{|\psi(L)|^2}{|\psi_{\pm}|^2}=(e^{i \gamma_B^{\pm} })^{b L/\pi},
\end{equation} 

\begin{figure}[h]
\centering
\includegraphics[width=0.5\textwidth]{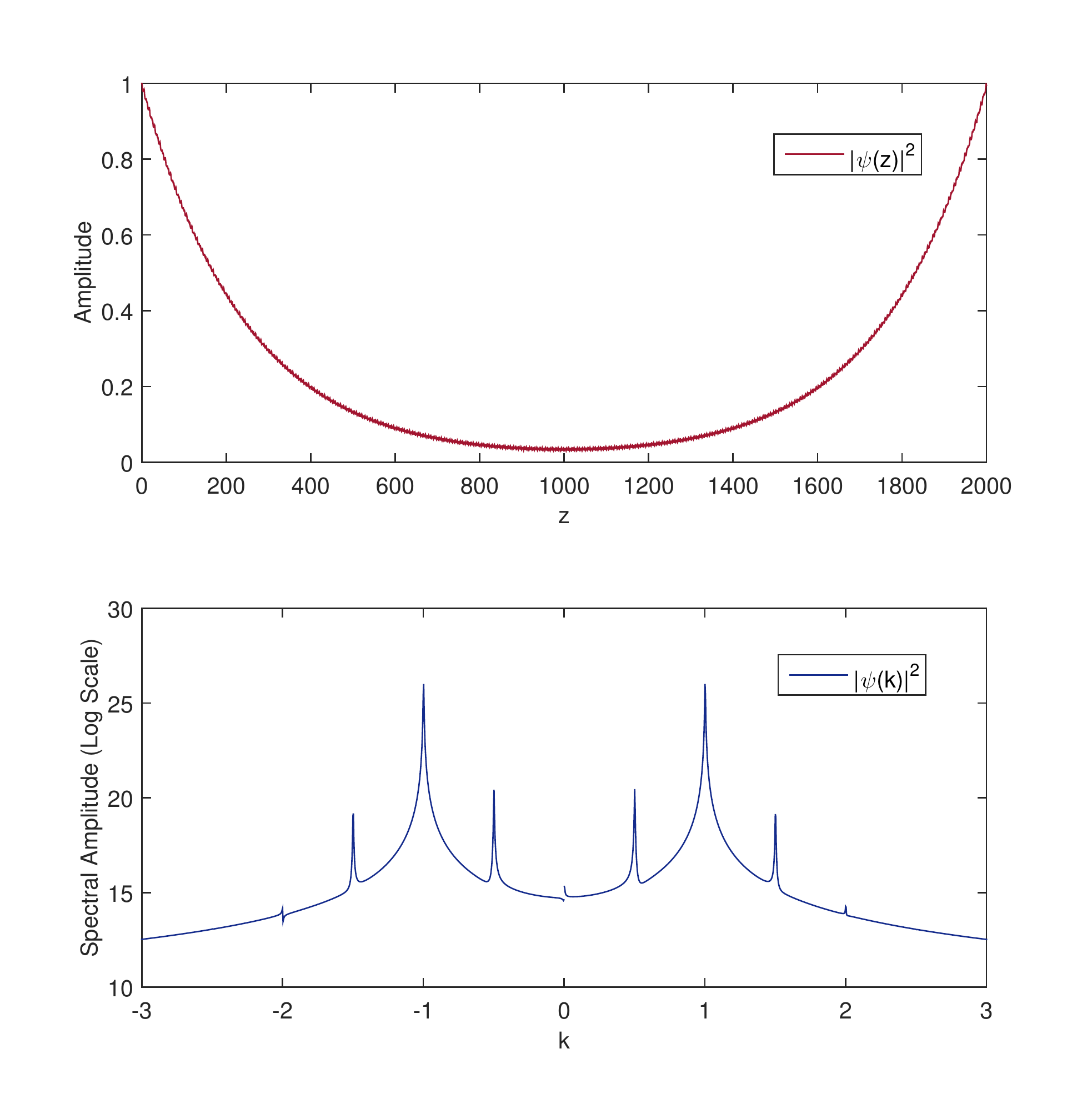}
\caption{a) The variation of the amplitude of the wavefunction with distance $z$ along the coupled waveguides, when the initial eigenstate is $\ket{\psi_+}$, for $a=0.12235, b=0.5$. b) The corresponding spectrum. Here, the peaks located at approximately $\pm1$ correspond to each instantaneous eigenstate. All other peaks are Floquet sidebands which are enhanced by the presence of the periodically varying geometric multiplier \cite{RosieandFabio}. The presence of both peaks, and the sudden conversion from loss to gain in $z$-space, indicates that the system is not evolving adiabatically.   }
\label{chargeswitch}
\end{figure}
where $b$ is the frequency of the parameter variation along $z$, $L$ is the total length of the coupled waveguides along the $z$-axis, and $|\psi(L)|^2$ is the amplitude of the evolved wavefunction at $z=L$. The formula comes from estimating how many complete cycles occur for distance $L$ ( If the function $\gamma_B(z)$ has a periodicity $2\pi$, a cycle will happen every $z=2\pi/b$, and hence the total number of cycles is $Lb/2\pi$). The function $i \gamma_B(z)$ is plotted in Figure \ref{berry} for a circular loop, a tilted loop, and a loop on the side of the monopole, like those visible in Figure \ref{figloops}. It is clear that varying the system's parameters such that a loop is traced out on the side on the monopole will induce net gain or loss in the coupled waveguide system.

\textit{Simulation results.}--- A loop on the side of the energy sheet of our system, such as that traced by the solid line in Figure \ref{figloops}, has the following parameterisation: $\kappa_R(z)=\cosh{[a \cos{(b z)}]}\cos{[-a \sin{(b z)}]}$, $\kappa_I(z)=\cosh{[a \cos{(b z)}]}\sin{[-a \sin{(b z)}]}$, $\gamma(z)=\sinh{[a \cos{(b z)}]}$, where $a$ controls the size of the loop, and $b$ is the frequency of the parameter variation along $z$. Figure \ref{gainandloss} a) shows the amplitude of a wavefunction as it evolves along the coupled waveguides for $a=0.1$ and $b=0.1$ when the plus instantaneous eigenstate is initially excited; the same is shown in c) for the minus eigenstate. The corresponding final spectra at the end of the propagation for a) and c) are displayed in b) and d) respectively. In each case, the dominant peak corresponds to the initially excited eigenstate, and a small second peak corresponding to the alternate eigenstate is present due to non-adiabatic effects. Floquet side bands are also present due to the periodic parameter variation, which can be intensified due to the modulation of the states' amplitude from the geometric multiplier, when it is itself periodic \cite{RosieandFabio}. If we calculate the factor in Equation (\ref{multiplier}) for this loop, for $|\psi_{\pm}|^2=0.5$ as seen in a) and c), we find $|\psi(L)|^2=0.36742$ for the plus eigenstate, and $|\psi(L)|^2=2.72168$ for the minus, which closely matches the final amplitude seen in a) and c) respectively. This confirms that the gain and loss pictured in Figure \ref{gainandloss} are due to the geometric multiplier, and the flux of the corresponding monopole through the chosen loop. In this case, as the loop is anticlockwise, we can expect a positive flux for the monopole, and negative for the antimonopole, leading to loss and gain respectively, and hence an unstable behaviour of the system.

In Figure \ref{chargeswitch}, we see the simulation results for the same anticlockwise side loop, with $a=0.12335$ and $b=0.5$, when the plus eigenstate is initially excited. These parameters are chosen as they will cause a seemingly perfect conversion of loss to gain for coupled waveguides of length $L=2000$. The peaks in the spectrum suggest an equally strong presence of the plus eigenstate and the minus eigenstate in the coupled waveguide system. If the system were evolving adiabatically, we would expect to see net loss, as in Figure \ref{gainandloss} a), and if we had initially excited the minus eigenstate, we would expect to see net gain, as in Figure \ref{gainandloss} b). Hence, what we appear to see is the system effectively switches from the dominant presence of one eigenstate (plus) to another (minus), and consequently a switch from the existence of the monopole in parameter space, to the existence of the antimonopole in parameter space: an analogue of charge reversal. 

Our claim of an effective state-flip is not extreme, as it was recently experimentally demonstrated that a waveguide which causes the parameters of a system to encircle an exceptional point can be used as an asymmetric mode switch; the encircling of the exceptional points causes the switch between different waveguide modes \cite{Rotter}. In our work, one could assume the breakdown of adiabaticity has led to a similar effect. If this coincides with the breaking of $\mathcal{PT}$ symmetry, then our one-sheet hyperboloid pictured in Figure \ref{figloops} will split into a two-sheet hyperboloid, as seen in \cite{Nesterov}.

\textit{Conclusions}--- We have shown that the presence of a non-periodic geometric multiplier in a non-Hermitian coupled waveguide system with periodically varying parameters and $\mathcal{PT}$ symmetry can induce net gain or loss along the waveguides. It is possible in such cases to induce a ``charge switch", where there is an apparent flip from one eigenstate to another in the system, which consequently implies a sign flip of the Berry curvature, analogous to a charge switch from a monopole to an antimonopole.  The first direct observation of Berry curvature effects in an optical system was recently made, which demonstrates that gaining experimental evidence for the existence of our monopole-antimonopole switch is possible \cite{Price2}.

\section{Acknowledgements}
The authors would like to acknowledge funding from the EPSRC Centre of Doctoral Training for Condensed Matter Physics (CM-CDT), grant number EP/L015110/1.

\end{document}